\documentclass[amsmath,10pt,nofootinbib,prl,superscriptaddress,twocolumn,prl,bibliography]{revtex4-1}
\usepackage{siunitx} 
\usepackage{microtype}
\usepackage{amsmath}
\usepackage{amsbsy}
\usepackage{amssymb}
\usepackage{graphicx}
\usepackage{color}
\usepackage{physics}
\usepackage{soul}
\usepackage{bm}
\usepackage{array}
\usepackage{multirow}
\usepackage{lipsum}
\usepackage{float}
\usepackage{upgreek}
\usepackage{hyperref}
\usepackage[capitalize]{cleveref}
\usepackage[normalem]{ulem} 
\usepackage{xr}
\usepackage[caption=false]{subfig}
\usepackage{microtype}
\usepackage{xcolor}

\usepackage{cancel}
\newcommand\redout{\bgroup\markoverwith
{\textcolor{red}{\rule[0.5ex]{2pt}{0.8pt}}}\ULon}

\makeatletter
\newcommand*{\addFileDependency}[1]{
  \typeout{(#1)}
  \@addtofilelist{#1}
  \IfFileExists{#1}{}{\typeout{No file #1.}}
}
\makeatother

\usepackage{verbatim}
\usepackage[bottom]{footmisc}
\makeatletter
\newcommand{\quickwordcount}[1]{%
  \immediate\write18{texcount -1 -sum -merge #1.tex > #1-words}%
  \immediate\openin\somefile=#1-words%
  \read\somefile to \@@localdummy%
  \immediate\closein\somefile%
  \setcounter{wordcounter}{\@@localdummy}%
  \@@localdummy%
}
\makeatother
\usepackage{natbib}

\begin{document}

\title{Interactions from memory footprints: Cage-hopping, mnemotaxis \& nonreciprocity}

\author{Jyoti Prasad Banerjee$^*$}
\affiliation{National Centre for Biological Sciences, Tata Institute of Fundamental Research,
Bengaluru, India}
\affiliation{School of Chemical and Biotechnology, SASTRA University, Thanjavur, India}

\author{Sambodh Sinha$^*$}
\affiliation{Department of Biological Sciences, Vanderbilt University, Nashville, TN, USA}
\author{Shashi Thutupalli}
\affiliation{National Centre for Biological Sciences, Tata Institute of Fundamental Research,
Bengaluru, India}
\affiliation{International Centre for Theoretical Sciences, Tata Institute of Fundamental Research, Bengaluru, India}

\author{Rituparno Mandal$^\dagger$}
\affiliation{James Franck Institute, University of Chicago, Chicago, IL 60637}
\affiliation{Soft Condensed Matter Group, Raman Research Institute, Bengaluru, India}

\author{Deb Sankar Banerjee$^\dagger$}
\affiliation{James Franck Institute, University of Chicago, Chicago, IL 60637}
\affiliation{Department of Chemical and Biological Sciences \& Department of Physics of Complex Systems, S. N. Bose National Centre for Basic Sciences, Kolkata, India}

\begin{abstract}
 
Cellular motility is integral to biological processes ranging from foraging to immune responses and cancer metastasis. While cells often navigate by sensing pre-existing environmental cues, many actively steer their future trajectories by remodeling their surroundings via shared, long-lived physicochemical trails called \emph{memory footprints}. These stigmergic responses to shared footprints also influence large-scale self-organization, such as the early stages of microcolony formation. Despite recent progress, a unified framework connecting single-cell dynamics and collective organization through effective pairwise interactions remains elusive.
To bridge this gap, we perform extensive agent-based simulations of an interacting random walk model that explicitly accounts for footprint deposition and decay. Our approach reveals distinct cage-hopping dynamics in solitary cells, as well as complex mnemotactic responses to externally imposed memory gradients. Furthermore, we demonstrate that shared spatial memory fields generate nontrivial effective pairwise interactions, both reciprocal and nonreciprocal, that shape the collective organization of cell populations. Beyond multiscale cellular dynamics, our findings have broader implications for the collective behavior of synthetic active matter, such as swarm robots.

\end{abstract}


\maketitle

\begingroup\renewcommand\thefootnote{$^*$}
\footnotetext{These authors contributed equally to this work.}
\endgroup
\begingroup\renewcommand\thefootnote{$^\dagger$}
\footnotetext{Email: rituparno$@$rri.res.in, debsbanerjee$@$bose.res.in}
\endgroup

\section*{Introduction}

Interactions via mediating fields appear in a wide variety of complex systems, such as plasmas~\cite{Schweigert1996}, soft matter~\cite{Likos2001}, and biological systems~\cite{Cherstvy2011}. For instance, the interaction between dusty plasma particles is mediated by a surrounding charge cloud~\cite{Shukla2009}. Such dust particles acquire a net negative charge from the plasma (due to the higher mobility of electrons), resulting in a screened electrostatic interaction typically modeled by the Yukawa potential~\cite{Thomas1994}. Similarly, interactions in soft and biological matter are frequently mediated by environmental degrees of freedom, such as solvent fluctuations~\cite{Lum1999}, elastic deformations~\cite{Yuval2013, Gupta2022}, or chemical fields~\cite{Pohl2014, Saha2014}. Notable examples in soft matter include depletion interactions (driven by entropic effects) between colloids in polymer solutions~\cite{AsakuraOosawa1954, AsakuraOosawa1958, Israelachvili2011}, fluctuation-driven attractive or repulsive Casimir forces between colloids~\cite{RevModPhys2018} or membranes~\cite{Sethna2012}, and oil-exchange-mediated droplet-droplet interactions~\cite{Meredith2020}.
\\
In biological systems, complex interactions mediated via shared physicochemical cues or fields span an extraordinary range of spatiotemporal scales. These interactions drive a remarkably diverse array of physiological and pathological processes, such as foraging~\cite{Czaczkes2015}, morphogenesis~\cite{Stookevaughan2018}, immune responses~\cite{Ryan2024}, and cancer metastasis~\cite{Elgundi2020}. For example, bridging-induced entropic attraction dictates the structural compaction of DNA~\cite{Brackley2013}; proteins interact through ions and solvent~\cite{Levy2006}; bacteria coordinate collective behaviors through quorum-sensing molecules~\cite{Mukherjee2019} and form microcolonies via self-deposited exopolysaccharide trails~\cite{Zhao2013}; and social insects organize through pheromone-mediated communication~\cite{Sumpter2006}. Similar principles also extend to ecological scales, where effective inter-species couplings govern community stability and biodiversity in ecological models~\cite{Volkov2007, Volkov2009}.
\begin{figure*}
    \centering   \includegraphics[width=\textwidth]{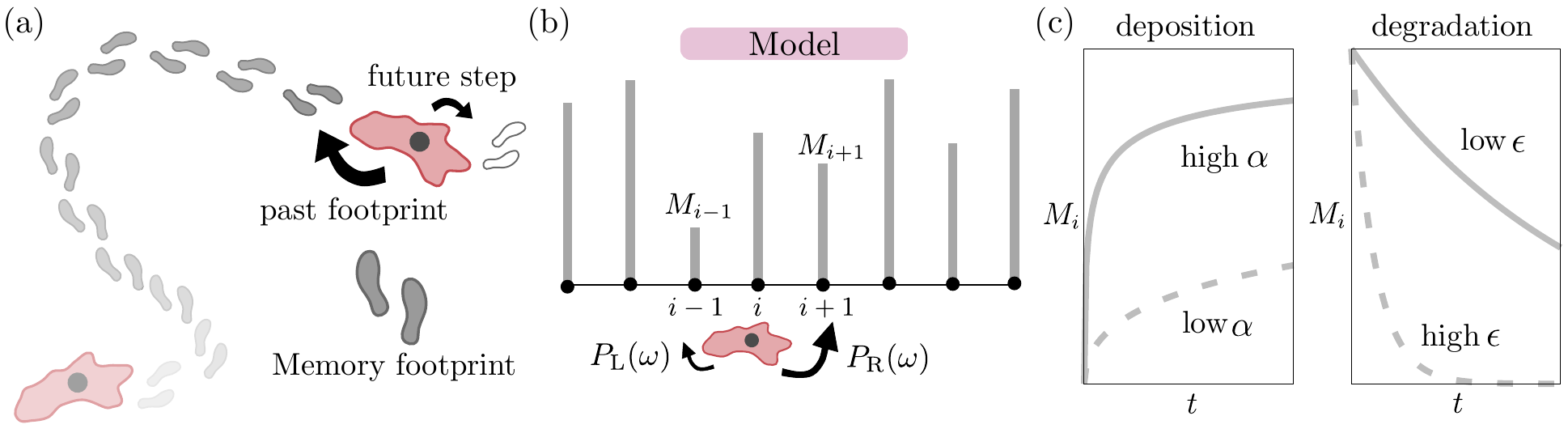}
    \caption{\textbf{Memory-mediated motility in a model cell.}
    (a) Cartoon of a cell leaving memory-footprints on the substrate during migration. Deposited footprints can couple a cell's current dynamics not only to its own past motion but also with that of other cells. (b) A schematic representation of the model: the cell is modeled as a biased random walker on a uniform 1D lattice. Its transition probabilities to the right \& left ($p_R(\omega)$ \&  $p_L(\omega)$, respectively) are biased by the memory-content of those neighboring sites and by the cell's response to memory, parametrized by $\omega$.
    (c) Memory deposition and degradation at the $i^{\text{th}}$ site: (left) a sketch showing how the memory-content of a site increases with time for high and low deposition coefficients $\alpha$ (assuming that the number of visits to that site, $n_i(t) \sim \sqrt{t}$, is the same for both cases at early times). (right) Decay of the memory content of a site with time for high and low degradation rates $\epsilon$.}
    \label{fig:1}
\end{figure*}
\\
 Among these, some are mediated by physical fields and their fluctuations, whereas some others arise active sensing of physicochemical cues in the local environment. Physically mediated interactions can occur between transmembrane proteins when they deform the surrounding membrane bilayer, generating mechanical strain fields. These strain fields subsequently drive long-range effective interactions~\cite{Goulian1993, Jiang2022} between such embedded proteins. On the other hand, sensing and responding to physicochemical cues help various social organisms navigate complex environments and achieve collective goals. Starving social amoebas (\emph{D. discoideum}) use extracellular cAMP gradients to initiate streaming or group migration~\cite{Artemenko2014}, trail-pheromones guide foraging ants and regulate colony recruitment~\cite{Czaczkes2015}, construction-pheromones help worker termites build intricate mounds~\cite{Ocko2019}, and mucus-bound chemical tracks allow gastropods to self-organize for mate-finding, communal homing, and desiccation prevention~\cite{Ng2013}.
\\
Sensing and responding to environmental cues (physicochemical fields) is especially crucial in cellular motility. Motility is often guided by pre-existing external physicochemical fields, in particular their gradients. Examples include mammalian leukocytes moving towards chemokines~\cite{Parent1999}, or fibroblasts migrating from a softer to a stiffer substrate~\cite{Lo2000}. However, some cells remodel and respond to their local environment via self-deposited cues, known as \emph{memory footprints}, and navigate without any pre-existing gradients. For instance, single epithelial cells deposit long-lived fibronectin and laminin footprints~\cite{d’Alessandro2021} that bias their future movements towards already visited regions. Memory produced by mechanical deformation can also orchestrate migration. As carcinoma cell clusters deform their viscoelastic collagen substrate, its slow relaxation generates a transient, asymmetric wake of high density and stiffness. Surfing away from this self-generated footprint enables persistent forward migration without external gradients or intrinsic polarity~\cite{Clark2022}.
\\
Interactions through self-generated memory-fields can give rise to rich single-cell dynamics: footprints deposited by solitary epithelial MDCK cells biochemically couple to the cell's polarity dynamics by triggering local integrin-mediated Rac1 activation~\cite{Perez2024}, which drives actin polymerization at the leading edge. This coupling between deposition and dynamics results in a transition between confined and exploratory modes of motility in both 1D and 2D. Stigmergy, the collective response to a shared memory field, also leads to complex many-body organization at larger scales. During biofilm initiation, surface-attached motile \textit{Pseudomonas aeruginosa} bacteria deposit Psl exopolysaccharides on the substrate. This strongly biases the twitching motility of the subsequent cells towards the heavily deposited regions, enabling the onset of heterogeneous surface colonization and microcolony formation~\cite{Zhao2013}.
\begin{figure*}
    \centering   \includegraphics[width=\textwidth]{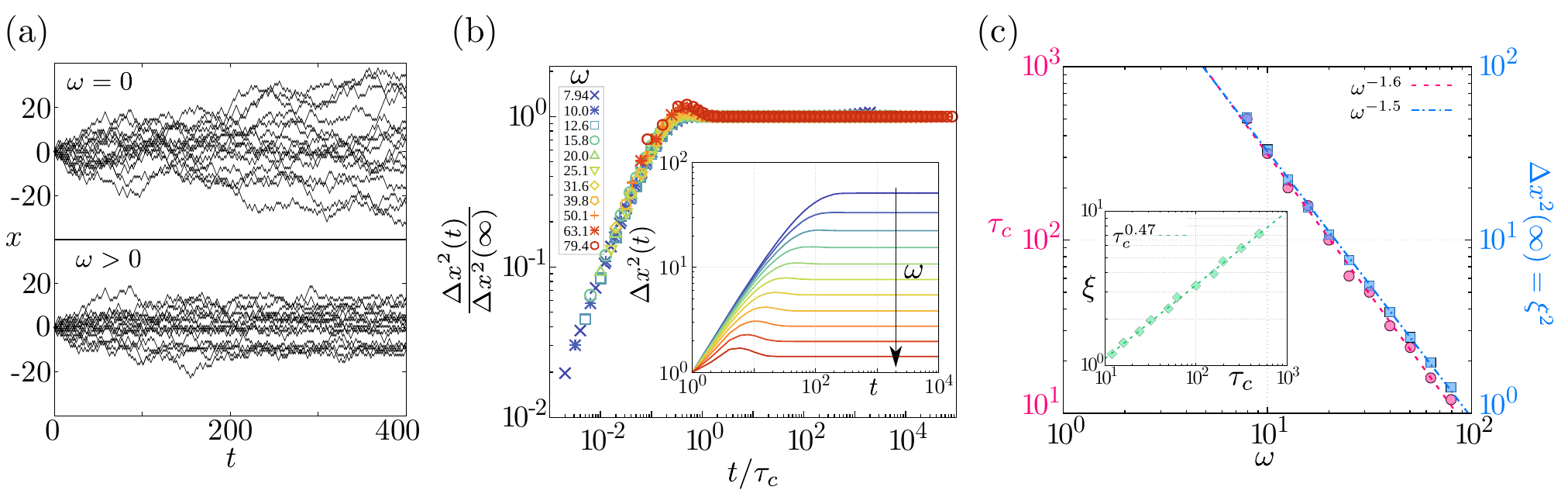}
    \caption{\textbf{Memory-induced self-caging.}
    (a) Representative trajectories of a solitary model cell in the absence (above) and presence (below) of a finite memory response coefficient, $\omega$. The trajectories above correspond to a pure random walk with the envelope spreading linearly with time. For $\omega \ne 0$ (below) the envelope stops spreading after some time, indicating self-caging. 
    (b) The mean squared displacement (MSD) of a cell from its origin  first increases diffusively, then saturates (a signature of self-caging). The inset shows that as $\omega$ increases (for fixed $\alpha$), the saturation occurs at smaller times and distances from the origin. When rescaled by the squared lengthscale ($\Delta x^2 (t \to \infty)=\xi^2$) and timescale ($\tau_c$), the MSDs collapse onto a single curve, as shown in the main panel.
    (c) Both $\tau_c$ (red) and $\xi^2$ (blue) decrease as power-laws in $\omega$, with slightly different exponents ($\tau_c \sim \omega^{-1.6}$, $\xi^2 \sim \omega^{-1.5}$). The inset shows the weakly sub-diffusive growth of $\xi$ with $\tau_c$, $\xi \sim \tau_c^{0.47}$. All power-law fits are shown in dashed and dash-dotted lines. We use $\alpha = 10^{-2}$ for all the panels in this figure.}     
    \label{fig:2}
\end{figure*}
\\
While most of the previous theoretical and computational studies have focused either on memory-mediated effective dynamics at the single organism level~\cite{Perez2024, Kranz2016}, or on organization and dynamics at collective scales~\cite{Agudo-Canalejo2019, Chatterjee2021, Uçar2025}, a generalized framework that connects single-cell dynamics and collective organization through effective pairwise interactions between two cells remains elusive.  
Here, we develop a model for cell motility using a collection of random walkers that do not interact directly but couple only through a shared field (memory footprints). We explicitly account for the deposition and degradation of the shared field and for their effect on cell dynamics.
For a single cell, we explore the role of memory response in self-caging, which turns into cage-hopping with the decay of memory. We then show that single cells with finite deposition ability respond in a complex manner to a pre-existing memory gradient. We subsequently explore the emergent effective forces, both reciprocal and nonreciprocal, between two cells interacting via a shared memory field. Finally, we investigate many-body organization for a specific reciprocal case and connect the single-cell, two-cell, and multicellular scales.

\section*{Model}\label{sec:model}

To understand how a cell interacts with its own memory footprint (see Fig.~\ref{fig:1}(a)) and how that, in the presence of a shared memory field, translates to memory-mediated effective interactions between cells, we develop a minimal model of cellular motility inspired by Self-attracting Random Walk (SATW) models~\cite{d’Alessandro2021}. 
For this, we consider individual cells as random walkers moving on a substrate represented by a lattice in 1D. The position of the $q^{\text{th}}$ cell is $x_q = i$ if it occupies the $i^{\text{th}}$ site of the lattice; the set of current positions of all the cells is denoted as $\{x_q\}$. These cells interact only via a shared memory field, $M_i$ (see Fig.~\ref{fig:1}(b)).
The probability of jumping from site $i$ to one of its nearest neighbor sites $j$ (where $j=i+1$ or $j=i-1$) is given by
\begin{equation}\label{eq:probabilities}
    p_j = \frac{e^{\omega_q M_j}}{Z}, \,\,\,\,\,\, Z = \sum_j e^{\omega_q M_j},
\end{equation}

\begin{figure*}
    \centering
    \includegraphics[width=\textwidth]{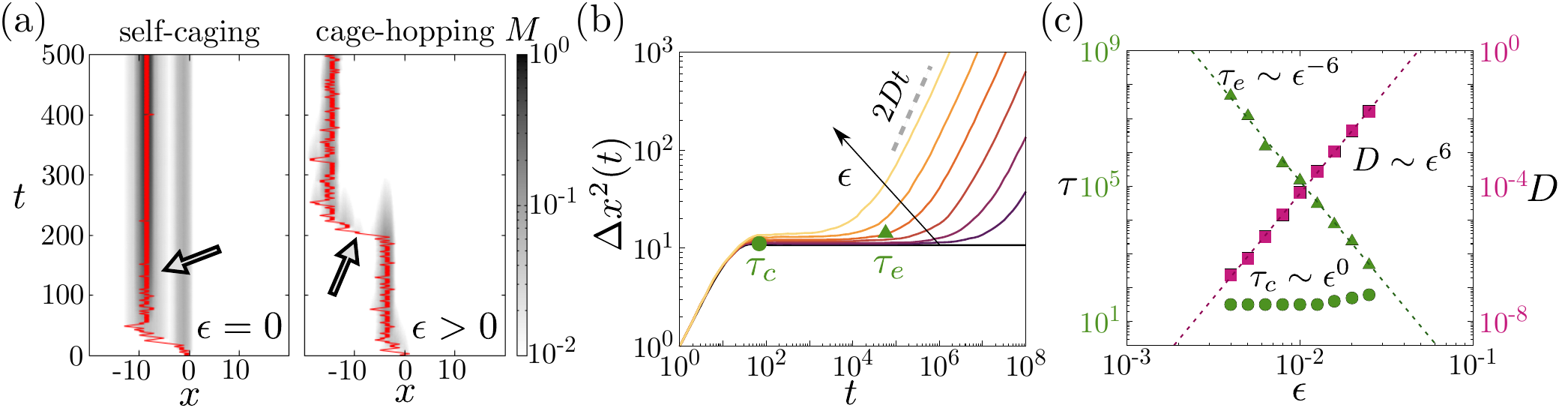}
    \caption{\textbf{Memory degradation leads to cage hopping.}
(a) Kymograph of a solitary cell (for $\alpha = 10^{-2}$, $\omega = 20$) showing self-caging ($\epsilon = 0$) and cage-hopping ($\epsilon > 0$). The memory-field is shown in grayscale in the background. For $\epsilon > 0$, one can notice the degradation of memory around the first cage ($x \sim -4$) right after the cell escapes ($t \sim 200$).
    (b) MSD from the origin, obtained from simulations (for $\alpha = 10^{-2}$, $\omega = 20$,  $\epsilon$ in the range $4 \times 10^{-3} - 3 \times 10^{-2}$), showing an initial diffusive regime, an intermediate caging regime, and a long-time diffusive regime. The case for $\epsilon = 0$  is shown in black for comparison.
    (c) While the caging timescale ($\tau_c$, green circles) weakly depends on the memory-degradation rate ($\epsilon$), the cage-escape timescale ($\tau_e$, green triangles) decreases with $\epsilon$ as a strong power law. The long-time diffusivity ($D$, magenta squares) also increases with $\epsilon$ as a strong power law, such that $D\tau_e$ remains constant.}
    \label{fig:3}
\end{figure*}

where $Z$ is the normalization factor. Here, $\omega_q$ (the subscript $q$ is omitted in single-cell cases for simplicity) is the memory response coefficient: a positive (negative) $\omega_q$ biases the cell toward the neighboring site with relatively higher (lower) memory content. The transition probabilities are constructed such that a cell cannot stay at its current position in the next time-step ($p_j = 0$ for $j = i$). The cell deposits memory at its current position. The deposited quantity depends on the cell's memory deposition coefficient ($\alpha_q$), the current memory content of that site, $M_i$, and the memory capacity of the substrate, $M_{\text{max}}$.
Memory is not permanent: it can degrade at a fixed rate at each site. These rules together lead to the following dynamical equation for the memory content at the $j^{\text{th}}$ site: 
\begin{equation}\label{eq:memory}
    \frac{dM_j(t)}{dt} = \frac{1}{M_{\text{max}}} \{M_{\text{max}} - M_j(t)\}^2 \sum_{i \in \{x_q\}} \alpha_q \delta_{ij} - \epsilon M_j(t).
\end{equation}
 Here the first term on the r.h.s. is the \emph{deposition} term with $\alpha_q$ (the subscript $q$ is omitted in single-cell cases for simplicity) as the memory deposition coefficient and $M_{\text{max}}$ as the memory capacity of the substrate (the rationale for this term is discussed in Appendix~\ref{app:modelRationale}). The second term is the linear \emph{degradation} term with $\epsilon$ as the memory degradation rate (see Fig.~\ref{fig:1}(c) for the effects of high and low values of $\alpha$ and $\epsilon$ on the memory content of the $i^{\text{th}}$ site). In the deposition term, the sum over $i \in \{x_q\}$ takes into account the current positions of all the cells. The absence of any direct steric interactions between cells allows for multiple occupancy of the lattice sites. Also note that in this study, the continuous-time Eq.~\eqref{eq:memory} is integrated with a unit time step, and the lattice spacing is unity.
 
 We consider $N \sim 1 - 10^3$ cells in this study. Typical results are averaged over $10^3-10^6$ independent realizations.
 We apply free boundary conditions in all simulations, except for the multicellular cases, which use periodic boundary conditions to emulate a ring geometry.
 We set $M_{\text{max}} = 1$ without loss of generality. 
 Memory deposition and response coefficients, $\alpha_q$ and $\omega_q$, can be thought of as cellular properties, while the memory degradation rate $\epsilon$ ($> 0$ for finite degradation) can be thought of as a property of the substrate.
 We restrict ourselves to $\alpha_q \geq 0$ and $\omega_q \geq 0$ throughout this work. Although the current work focuses on a specific region of the parameter space, our model is general and provides a framework for a much broader class of memory-mediated interactions. For example, cases with $\alpha_q < 0$ in the presence of an external supply of chemical cues provide an opportunity to study the effects of interactions through local depletion gradients, some of which have been observed through experiments and simulations in Ref.~\cite{Tweedy2020}. In the same way, $\omega_q < 0$ leads to self-avoidance and a plethora of collective phenomena as reported in Ref.~\cite{Hokmabad2022}. These cases, along with other possible extensions of the model, such as the introduction of persistence~\cite{d’Alessandro2021}, memory of a vectorial nature (such as trails of Pharaoh's ants providing polarity information~\cite{Jackson2004}), and the intricacies of effective cellular interactions in 2D, will be addressed in future work.

\section*{Results}\label{sec:results}

\subsection{Memory response leads to self-caging}\label{subsec:results_caging}

We first consider the simple case without memory degradation (\textit{i.e.}, we set $\epsilon = 0$). For a vanishing memory deposition rate ($\alpha=0$), a vanishing memory response coefficient ($\omega=0$), or both, our model cell behaves as a simple unbiased random walker (Fig.~\ref{fig:2}(a), top). However, as $\omega$ is increased while keeping $\alpha$ at a fixed finite value, the cell, enslaved by its own memory footprints, deviates from the simple random walk behavior, eventually becoming {\emph{self-caged}} within a few sites (Fig.~\ref{fig:2}(a), bottom). This self-caging behavior becomes stronger with increasing $\omega$. This behavior has been reported previously for both chemoattractants~\cite{Tsori2004, Kranz2016, BarbierChebbah2022, Khuntia_2022} and chemorepellents~\cite{Hokmabad2022}. To quantify this, we compute the mean squared displacement (MSD) from single-cell trajectories for different values of $\omega$ (see Fig.~\ref{fig:2}(b)). We find that increasing $\omega$ decreases the caging timescale $\tau_c$, which is the time the cell takes to become self-caged starting from the origin (defined as the onset time at which the MSD enters its plateau, see Fig.~\ref{fig:2}(b), inset). It also decreases the caging lengthscale, $\xi=\sqrt{\Delta x^2(\infty)}$ (where $\Delta x^2(\infty)$ is the MSD at large time $t\to \infty$), {\it{i.e.}}, the size of the confinement of the trajectories around the origin (Fig.~\ref{fig:2}(b), inset).
 The MSD curves obtained for different values of $\omega$ collapse onto a single master curve when rescaled by the caging lengthscale ($\xi$) and the caging timescale $\tau_c$ (see the main panel of Fig.~\ref{fig:2}(b)).
Next, we study how this emergent caging lengthscale ($\xi$) and timescale ($\tau_c$) change with the memory response coefficient $\omega$. We find that both follow power-law scaling, $\xi^2 \sim \omega^{-1.5}$ and $\tau_c \sim \omega^{-1.6}$ (Fig.~\ref{fig:2}(c)). Consequently, $\xi \sim \tau_c^{0.47}$ (Fig.~\ref{fig:2}(c), inset): the caging lengthscale grows weakly subdiffusively with the caging timescale, which we attribute to memory-mediated self-correlated (self-attractive) dynamics.
\begin{figure*}
    \centering
    \includegraphics[width=\textwidth]{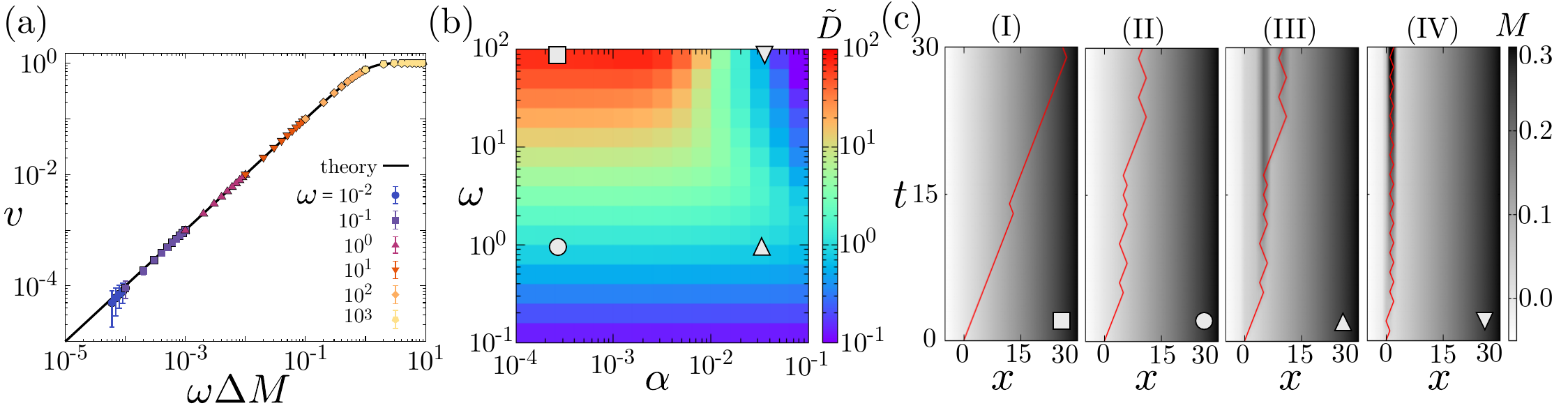}
    \caption{\textbf{Response to pre-existing memory-gradient.} 
    (a) The coarse-grained velocity $v$ (measured using displacement over a 100-step window) for $\alpha = 0$, $\omega = 10^{-2}$--$10^3$ (different symbols and colors) and $\Delta M = 10^{-3} - 10^{-2}$, matches the exact analytical prediction (black solid line). For $\omega \Delta M \ll 1$, the cell's mobility (effective response) to the pre-existing memory-gradient, $\tilde{D} = v/\Delta M \approx \omega$. The error bars represent the standard error of mean. 
    (b) The phase diagram of mobility, $\tilde{D}$, in the $\alpha\text{-}\omega$ plane shows that for $\alpha > 0$, $\tilde{D}$ deviates from $\omega$, and becomes nonmonotonic at large $\alpha$. 
    (c) Typical kymographs (red: cell trajectory; grayscale: memory field) from four representative points in the $\alpha\text{-}\omega$ plane, marked by open symbols (square, circle, upward triangle, and downward triangle).}
    \label{fig:4}
\end{figure*}

\subsection{Memory decay facilitates cage hopping}

In this section, we introduce the decay of deposited memory footprints over time. This decay may arise from chemical degradation, deactivation, or unbinding of deposited molecules such as fibronectin and laminin~\cite{d’Alessandro2021, lu2011extracellular}, or the relaxation of mechanical fields such as the density~\cite{Clark2022} or strain field~\cite{humphrey2014mechanotransduction, van2016strain} that carry a signature of memory footprints. 
We explore the effect of a finite memory degradation rate ($\epsilon > 0$) on the dynamical behavior of a solitary cell. We observe a new feature in single-cell trajectories: {\it cage-hopping}. Comparing the kymographs of a solitary cell without (left) and with (right) memory degradation shows that, with degradation, the cell keeps hopping between successive caged episodes (Fig.~\ref{fig:3}(a)). This is reminiscent of the cage-hopping events, a characteristic of glassy dynamics in supercooled liquids~\cite{Weeks2000, Berthier2011}. However, in our model such cage-hopping is a single-particle phenomenon, rather than a collective process, as observed in glassy systems. In contexts of both cellular migration and animal foraging, cage-hopping dynamics may act as a physical realization of intermittent search strategies~\cite{Benichou2011} and area-restricted search~\cite{Dorfman2022}, offering distinct advantages in crowded, patchy, or heterogeneous environments\cite{Vergassola2007,Godec2015}.
\\
A statistical analysis of the trajectories confirms this behavior: the MSD shows an intermediate-time plateau and a long-time diffusive regime (see Fig.~\ref{fig:3}(b)).
A new timescale emerges here: the cage-escape timescale $\tau_e$. We find that $\tau_e$ and the long-time diffusivity ($D_\infty$) depend strongly on the memory degradation rate $\epsilon$. With increasing degradation rate, the cage-escape timescale decreases while the long-time diffusivity increases (Fig.~\ref{fig:3}(b), (c)), both following power laws. 
A few observations are worth noting here (Fig.~\ref{fig:3}(c)): (1) the caging timescale $\tau_c$ is weakly dependent on the memory degradation rate $\epsilon$ and is controlled mainly by the memory-response coefficient, $\omega$, as shown in Fig.~\ref{fig:2}(c);
(2) Curiously, $D_\infty$ and $\tau_e$ show strong dependence on $\epsilon$, with $D_\infty \sim \epsilon^{6}$ and $\tau_e \sim \epsilon^{-6}$; and
(3) $D_\infty \propto \tau_e^{-1}$, akin to the Stokes--Einstein relation $D_{\infty} \tau_{\alpha} = \text{const.}$ between the long-time diffusivity $D_{\infty}$ and the density-relaxation time ($\tau_{\alpha}$) in supercooled liquids. However, the extent of this analogy with glassy dynamics merits further exploration beyond the scope of the present study.

\subsection{Mnemotaxis}

\begin{figure*}
    \centering
    \includegraphics[width=\textwidth]{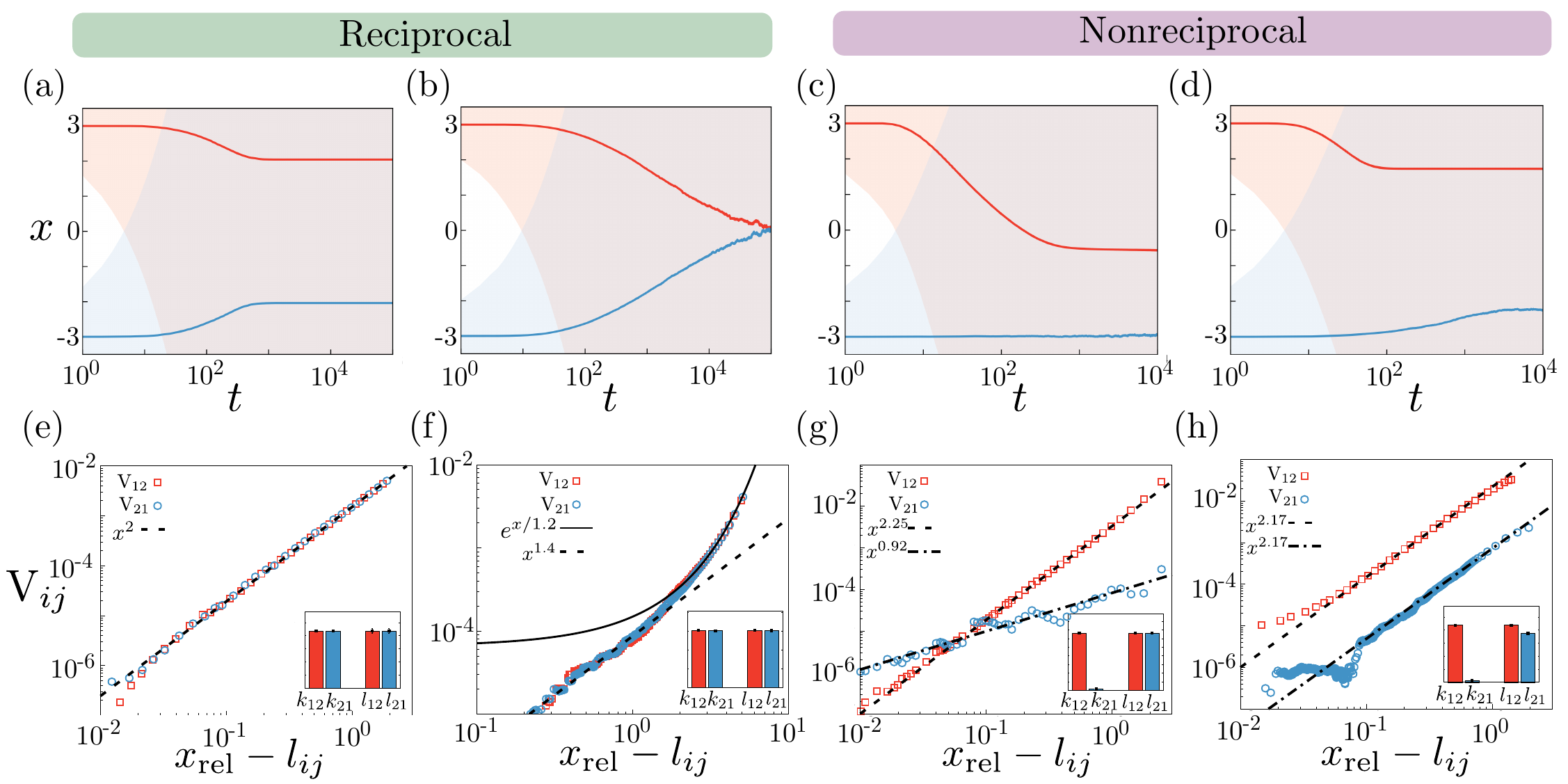}
    \caption{\textbf{Reciprocal and nonreciprocal effective pairwise interactions.} 
    (a)--(d) Averaged trajectories of two cells (red and blue solid lines) with the initial separation $x_0 = 6$. The shaded regions indicate the standard deviation over $10^6$ independent runs. (e)--(h) The corresponding potentials $V_{ij}$ of the form $V_{ij} = \frac{k_{ij}}{2} \left( x_{\text{rel}} - l_{ij}\right)^{\delta_{ij}}$, where the fit parameters ($k_{ij}$ and $l_{ij}$) are shown as relative strengths in the inset bar plots; the fitted exponents $\delta_{ij}$ are given in the legends and in Table~\ref{tab:inferred_parameters}. 
(a), (e) Reciprocal type I: $\alpha_1 = \alpha_2 = 3 \times 10^{-4}, \omega_1 = \omega_2 = 10^2$. One-sided harmonic attractive effective potential. The cells approach each other with an effective harmonic attraction, but eventually get caged at a finite separation. 
(b), (f) Reciprocal type II: $\alpha_1 = \alpha_2 = 2 \times 10^{-2}, \omega_1 = \omega_2 = 1.5$. The cells approach and eventually meet each other. The potential is initially exponential and then crosses over to a power law with an exponent smaller than 2 ($\delta \approx 1.4$).
(c), (g) Nonreciprocal type I: $\alpha_1 = 3 \times 10^{-4}, \alpha_2 = 2 \times 10^{-2}, \omega_1 = 10^2, \omega_2 = 1.5$. The 1st cell approaches the 2nd one, while the 2nd one is only slightly attracted to the 1st one. $\delta_{21} < \delta_{12}$, $k_{21} \ll k_{12}$.
    (d), (h) Nonreciprocal type II: $\alpha_1 = 3.98 \times 10^{-3}, \alpha_2 = 1 \times 10^{-2}, \omega_1 = 3.98 \times 10^1, \omega_2 = 2.51$. Here the 1st cell approaches the 2nd one early, while the 2nd one approaches the 1st later.} 
    \label{fig:5}
\end{figure*}

Before turning to the effective interactions between two cells interacting via a shared memory field, it is crucial to understand the dynamical behavior of a single cell in the presence of a preexisting, externally imposed spatial memory gradient. Because cell movement here is guided by memory footprints, we call it {\it mnemotaxis} (\textit{mnemo-}, from the Greek for ``memory''). 
First, we consider the simple case of mnemotaxis in which the cell can only respond to memory ($\omega > 0$) but cannot deposit memory ($\alpha=0$) and moves in an externally imposed, spatiotemporally invariant memory gradient (\textit{i.e.}, $\Delta M=M_{i+1}-M_i$ is the same for any $i$ at any time). In this case, the cell necessarily follows the imposed gradient, since $\omega > 0$.

The coarse-grained velocity, $v$, of the cell, defined as the ratio of its displacement to the elapsed time, can be computed directly (see Appendix~\ref{app:mnemotaxisDerivation} for details) from Eq.~\eqref{eq:probabilities} as
\begin{equation}\label{eq:velocity}
    \begin{split}
        v & = p_R - p_L = \tanh{(\omega \Delta M)}
    \end{split}
\end{equation}
where $p_R$ and $p_L$ are the transition probabilities to jump to the right and left neighboring site, respectively ($v$ is measured in lattice spacings per time step). The velocity measured in simulations for various values of $\omega$ collapses onto the analytical prediction, Eq.~\eqref{eq:velocity} (see Fig.~\ref{fig:4}(a)).
 For $\omega \Delta M \ll 1$, {\it i.e.}, in the linear regime, the gradient response coefficient or mobility, $ \tilde{D} = v/\Delta M = \omega - \mathcal{O}(\omega^3)$. This linear dependence of $\tilde{D}$ on $\omega$ provides the baseline for the case of finite $\alpha$.

Once $\alpha$ is finite, the cell deposits memory on top of the externally imposed gradient and responds to the resulting memory profile. As $\alpha$ increases, the dependence of the mobility $\tilde{D}$ on $\omega$ departs significantly from linearity and becomes nonmonotonic (see Fig.~\ref{fig:4}(b)). 
To illustrate, we choose four points in the response diagram representing all four combinations of a small and a large value of $\alpha$ and $\omega$, marked by the four symbols in Fig.~\ref{fig:4}(b).
We obtain representative kymographs for the four specified sets of $\alpha$ and $\omega$ values against the background of the external memory gradient. This shows how the cells with different $\alpha$ and $\omega$ modify the memory profile, and how, in turn, the modified memory profile influences their motion (see Fig.~\ref{fig:4}(c)). These results demonstrate that the cell can have diverse responses based on its own ability to create and respond to the memory, even in an external memory gradient, such as one generated by neighboring cells. This suggests that cells sharing a memory field can experience memory-mediated effective interactions. We now focus on these effective interactions by exploring two-cell and many-cell scenarios in the next sections.

\subsection{Effective two-body interactions}

\begin{figure*}
    \centering
    \includegraphics[width=\textwidth]{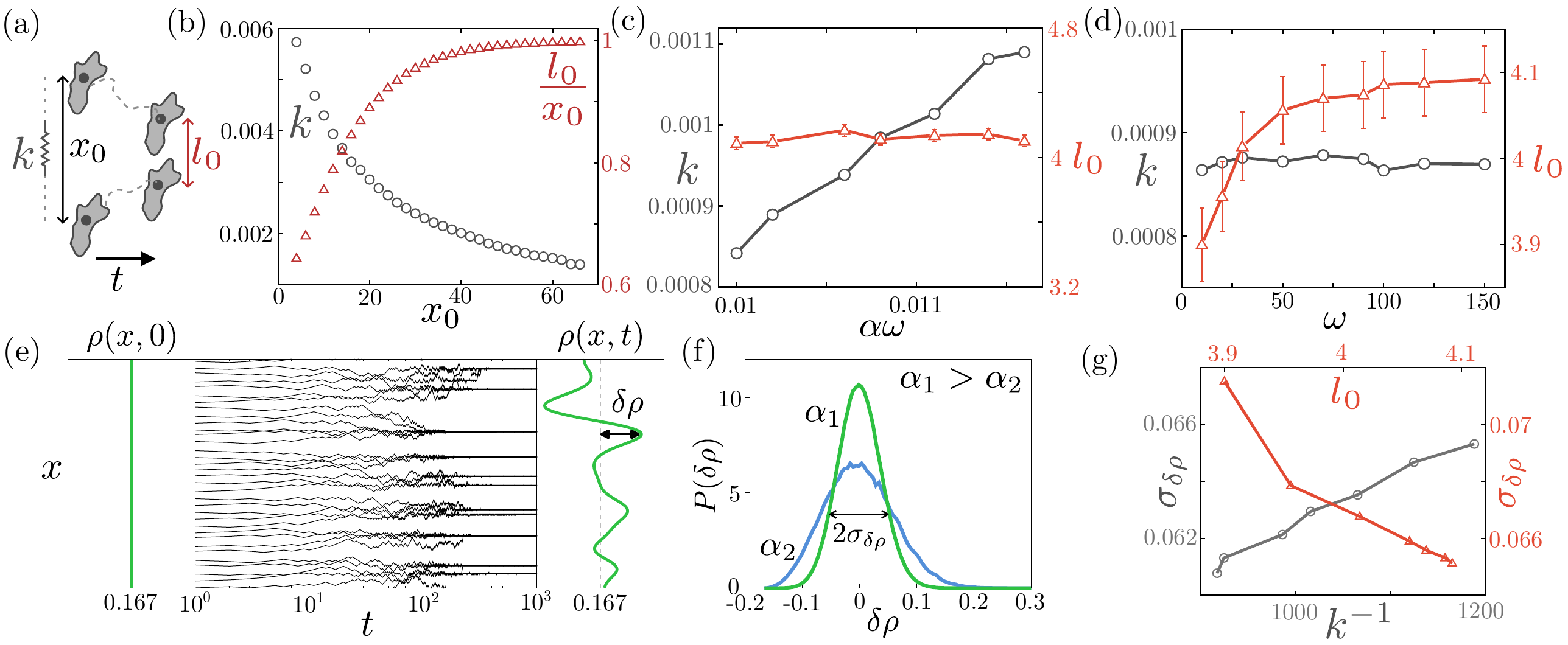}
    \caption{\textbf{Connecting single-cell properties, pairwise interactions, and collective organization.}
    (a) A schematic showing the interaction between two cells with parameters ($\alpha=3 \times 10^{-4}, \omega= 10^2$), as in Fig.\ref{fig:5}(a). 
    (b) Dependence of the spring constant $k$ (black circles) and the rescaled rest length $l_0/x_0$ (red triangles) on the initial separation $x_0$. 
    (c) Within the region in the $\alpha\text{-}\omega$ plane with harmonic-like effective pairwise interactions, increasing $\alpha \omega$ (keeping $\omega$ fixed) leads to an increase in $k$ while $l_0$ remains almost constant. 
    On the other hand, (d) increasing $\omega$ (keeping $\alpha \omega$ fixed) increases $l_0$ while $k$ remains nearly unchanged. 
    (e) Trajectories of many ($N \sim 10^2$) identical cells (with $\alpha = 3 \times 10^{-4}, \,\,\, \omega = 10^2$), initiated with $x_0=6$ (leading to a uniform initial line-density of $ \rho(x, 0) = 1/6$). As time progresses, multiple ($2$ -- $7$) cells aggregate into nonmotile clusters, resulting in a heterogeneous density profile. 
    (f) Distribution of density variation, $\delta \rho (x,t) = \rho(x,t) - \rho(x,0)$, at long times, for $\alpha = \alpha_1 = 1.2 \times 10^{-4}$ (shown in blue) and $\alpha = \alpha_2 = 3 \times 10^{-4}$ (shown in green), keeping $\omega_i = 10^{2}$ fixed. 
    (g) The standard deviation of the $\delta\rho$-distribution, representing the inhomogeneity in the cell-clusters, is shown as a function of the two pairwise interaction parameters, $k$ (black, bottom axis showing $k^{-1}$) and $l_0$ (red, top axis).
    } 
\label{fig:6}    
\end{figure*}

In this section, we focus on effective interactions between two cells depositing a shared memory footprint. To study these interactions quantitatively, we select the parameter values of the two cells from the phase diagram in Fig.~\ref{fig:4}(b) and place them at an intercellular separation of $x_0$. We also set $\epsilon=0$ in this section; the case $\epsilon > 0$ is discussed in Appendix~\ref{app:twoBodyFiniteDecay}.
Initially, these two cells move independently and deposit memory in the neighboring regions of their initial positions. These regions grow over time; once they merge (the red and blue shaded areas in Fig.~\ref{fig:5} top row, which indicate the spread of the trajectories), the cells start interacting via their shared memory field.
\\
Although individual cell trajectories are stochastic for any given realization, they reveal a systematic trend when averaged over the ensemble.  
These averaged trajectories would be straight horizontal lines (similar to the initial part of the trajectories shown in   Fig.~\ref{fig:5} top row) in the absence of memory deposition ($\alpha=0$) or memory response ($\omega=0$), since each cell would then behave as a simple random walker, with no interaction between them. By varying the values of the memory deposition coefficient $\alpha$ and memory response coefficient $\omega$, we find qualitatively distinct types of emergent memory-mediated interactions between the two cells. We categorize the interaction as either \textit{reciprocal} where cell--$1$ influences cell--$2$ in the same way as cell--$2$ influences cell--$1$, or \textit{nonreciprocal}, where the two cells influence each other differently. This aspect of the reciprocity (nonreciprocity) is reflected in the symmetry (asymmetry) of the trajectories, as seen in Fig.~\ref{fig:5}, top row. It is noteworthy here that in recent years, nonreciprocal interactions, characterized by the breakdown of action-reaction symmetry, have emerged as a foundational paradigm in nonequilibrium and active matter physics~\cite{Soto2014, Ivlev2015, Fruchart2021, Bowick2022, Mandal2024}. 
\\
For a quantitative analysis, we estimate the interaction force on cell--$1$ due to cell--$2$, $F_{12}$, and vice versa, $F_{21}$, from the cell trajectories and infer the respective potentials $V_{12}$ and $V_{21}$ (see Appendix~\ref{app:twoBodyEffectivePotentials} for details). Note that for a reciprocal system $V_{12} = V_{21}$, whereas in the case of nonreciprocal interactions $V_{12} \neq V_{21}$. Our analysis indicates that the dependence of the potentials on the intercellular separation ($x_{\text{rel}} = x_j - x_i$) is well characterized by the following shifted power law. 
\begin{equation}
    V_{ij} = \frac{k_{ij}}{2} (x_\text{rel} - l_{ij})^{\delta_{ij}}
\label{eq:potential}    
\end{equation}
where $k_{ij}$, $l_{ij}$, and $\delta_{ij}$ are fit parameters of the numerically inferred potentials. For reference, if the exponent $\delta_{ij} = 2$, the potential resembles a harmonic one with $k_{ij}$ as the spring constant and $l_{ij}$ as the rest length.
\\
We find that the effective interaction is reciprocal (attractive) when $\alpha_1=\alpha_2$ and $\omega_1=\omega_2$, but the averaged trajectories and the inferred potentials depend on the particular parameter values. 
We refer to the cases of Fig.~\ref{fig:5}(a),(e) and Fig.~\ref{fig:5}(b),(f) as reciprocal type I and type II, respectively. The inferred potentials in these cases reflect this reciprocity, {\it{i.e.}}, $V_{12}=V_{21}$ (see Fig.~\ref{fig:5} (e),(f)) and the fits yield the same values of the parameters $k_{ij},\, l_{ij} \,$ and $\delta_{ij}$ for both cells (within fitting accuracy) (see Fig.~\ref{fig:5}(e), (f) inset). 
In some cases, the inferred potential is exponential at early times (large $x_\text{rel}$) and crosses over to a power law at later times (small $x_\text{rel}$) (see Fig.~\ref{fig:5}(f)).
\\
Nonreciprocity in the memory-mediated interactions emerges when $\alpha_1\neq \alpha_2$ and/or $\omega_1\neq \omega_2$ (see Fig.~\ref{fig:5}(c), (d), (g), (h); inferred potentials clearly show $V_{12}\neq V_{21}$). In the cases studied, the cell with higher $\omega$ and lower $\alpha$ ($\alpha_1 < \alpha_2$ \& $\omega_1 > \omega_2$) is attracted toward the other cell while the second cell shows only little attraction toward the first one (as observed in Fig.~\ref{fig:5}(c), (d)). We characterize such emergent nonreciprocity in two categories, type I and type II. In type I, the fitted parameters $k_{ij}$ and $\delta_{ij}$ differ significantly between the two cells while $l_{ij}$ remains the same, as shown in Fig.~\ref{fig:5}(g) and Table~\ref{tab:inferred_parameters}. Type-I behavior occurs for $\alpha_1 \omega_1 = \alpha_2 \omega_2$ (here $\alpha_1\omega_1=\alpha_2\omega_2=0.03$), in which case the trajectory of the 2nd cell (see Fig.~\ref{fig:5}(c); marked in blue) remains almost unaffected by the 1st cell. In contrast, $\alpha_1 \omega_1 \neq \alpha_2 \omega_2$ leads to type-II nonreciprocity. Here, although $k_{ij}$ and $l_{ij}$ differ significantly, the $\delta_{ij}$ are nearly equal (Table~\ref{tab:inferred_parameters}). Moreover, the 2nd cell responds only weakly and with a delay to the 1st cell (compare Fig.~\ref{fig:5}(d) with Fig.~\ref{fig:5}(a)), in stark contrast to the reciprocal cases. The details of the inferred parameters are given in Appendix~\ref{app:twoBodyInferredParam}.

\subsection{Bridging the scales}

In the previous section, we demonstrated the emergence of diverse types of effective interactions between two cells. Here we discuss how variations in the ability to produce and respond to the memory field at the single-cell level can give rise to such diversity and complexity. For this, we analyze quantitatively how the intercellular effective interaction changes with single-cell parameters (such as  $\alpha$, $\omega$, and the cell separation $x_0$). We consider the reciprocal case where the inferred potential is approximately harmonic with $\delta \approx 2$ (see Fig.~\ref{fig:5}(a),(e)) and redefine the potential parameters as $k=k_{12}=k_{21}$ and $l_0=l_{12}=l_{21}$ (see Fig.~\ref{fig:6}(a)). 
First, we vary the initial cell separation $x_0$ and find that the interaction strength $k$ decreases and the rest length $l_0$ increases with increasing $x_0$ (see Fig.~\ref{fig:6}(b)), indicating that the interaction weakens at large separations. We also find that $k$ and $l_0$ of the emergent interaction can be modulated approximately independently by varying the single-cell parameters in specific ways. While, for fixed $\omega$, the strength of the interaction increases with increasing $\alpha \omega$ (see Fig.~\ref{fig:6}(c)), the rest length ($l_0$) remains nearly constant. Conversely, at fixed $\alpha\omega$, the rest length $l_0$ increases with increasing $\omega$, whereas the interaction strength remains almost unchanged (see Fig.~\ref{fig:6}(d)). 
Next, we explore the collective behavior of many cells and rationalize it using the effective interactions developed above.

\subsection{Collective behavior} 
We now study the dynamics of many cells depositing memory as they move, again considering the reciprocal type-I case. We start with $10^3$ identical cells placed a distance ($x_0=6$) apart on a ring lattice (periodic boundary conditions). This initial condition corresponds to an initial coarse-grained cell density $\rho(x,0)=1/6$. 
As the cells move and deposit memory, the homogeneous state becomes unstable and the cell density $\rho(x,t)$ develops large variations in the steady state, quantified by $\delta\rho(x,t)=\rho(x,t) - \rho(x,0)$ (see Fig.~\ref{fig:6}(e)).
We characterize these variations by the standard deviation ($\sigma_{\delta\rho}$) of the distribution of density variations $P(\delta\rho)$, computed at late times. We find that $\sigma_{\delta\rho}$ decreases with increasing $\alpha$ at fixed $\omega$ (see Fig.~\ref{fig:6}(f)).
Next, we examine how the two-cell interaction strength $k$ and rest length $l_0$ modulate the density variations. Fig.~\ref{fig:6}(g) shows that $\sigma_{\delta\rho}$ decreases with both increasing $k$ (\textit{i.e.}, decreasing $k^{-1}$, which is proportional to the timescale of the effective two-cell interaction) and increasing $l_0$. However, $\sigma_{\delta \rho}$ is considerably more sensitive to changes in $l_0$ than to changes in $k$ ($\frac{\Delta\sigma_{\delta\rho}}{\Delta l_0} \approx -2.06$, whereas $\frac{\Delta\sigma_{\delta\rho}}{\Delta k} \approx -0.25$, where $\Delta$ represents the percentage change). These results demonstrate how the two-cell interaction parameters modulate the
 steady-state density fluctuations of many cells interacting through a shared memory footprint.

\section*{Discussion}\label{sec:summary}

While certain cells migrating across a substrate are known to leave physicochemical trails on the substrate, a fundamental question remains: how do self-deposited spatial memory footprints influence the cell's dynamics and mediate effective pairwise interactions between cells to dictate multi-scale self-organization? To address this, we have developed a minimal interacting random walk model that explicitly incorporates the deposition and degradation of memory-footprints, as well as the cell's responsiveness to them. Through extensive computational modeling we have shown that this framework bridges dynamical features of a single cell with effective pairwise interactions and population-level density fluctuations. 
\\
At the single-cell level, our results demonstrate that the memory response coefficient ($\omega$) induces localization or self-caging, consistent with prior studies~\cite{Tsori2004, Kranz2016, BarbierChebbah2022, Khuntia_2022}. When memory degradation is introduced ($\epsilon > 0$), this confinement becomes transient, leading to single-particle cage-hopping dynamics, reminiscent of the structural relaxation in glassy systems~\cite{Weeks2000, Berthier2011}. The long-time diffusivity in this case shows a strong power-law dependence on the degradation rate: $D_\infty \sim \epsilon^6$. In the presence of an external memory field, cells exhibit mnemotaxis, where the memory deposition coefficient ($\alpha$) nontrivially modulates the effective mobility ($\tilde{D}$). 
Building upon this, when two cells interact through a shared memory field, these single-cell properties can manifest as emergent reciprocal or nonreciprocal effective pairwise forces. Identical cells experience symmetric attractive potentials whose rest length and stiffness can be tuned independently via cellular deposition and response coefficients. Conversely, asymmetry in memory deposition ($\alpha_1 \neq \alpha_2$) or responsiveness ($\omega_1 \neq \omega_2$) breaks action--reaction symmetry, generating two types of emergent nonreciprocal dynamics (types I and II), in which the cells show unequal or nonmutual affinities toward each other. We note that the nonreciprocal effective interactions arising from mismatches of the single-cell parameters in our model may be realized even within a single type of cells in certain cancer cell populations. Compared to normal tissues, these cells often exhibit vastly broader spatial and intercellular heterogeneities in substrate remodeling (through deposition or deformation~\cite{Pickup2014,Kim2020}), and motility responses~\cite{Wirtz2011,Paul2017,Kwon2019}.
A recent work reports the existence of nonreciprocal interactions between cancer cells using data-driven inferences on experimental multicellular trajectories~\cite{Hoogland2026}. However, further investigations are needed to uncover whether memory-mediated interactions are crucial for such emergent nonreciprocal interactions between cells. 
Moving on to the collective scale, we find that these pairwise interactions destabilize the uniform state, causing the population to partition into nonmotile clusters with a stationary distribution of density fluctuations. We show a clear correlation between the standard deviation ($\sigma_{\delta \rho}$) of this distribution and the parameters ($k$, $l_0$) characterizing the effective pairwise interaction.
\\ 
We restricted our model to dynamics on a one-dimensional lattice. This framework directly applies to cellular migration within narrow microfluidic channels~\cite{Irimia2009}, quasi-1D extracellular matrix (ECM) fiber tracks, and patterned substrates in 1D~\cite{d’Alessandro2021}. Previous single-cell models incorporating cell polarity and persistent motion in 1D and 2D have reported rich oscillatory and circular motility modes~\cite{Perez2024}. While our current agents lack intrinsic polarity or persistence, extending our model to 2D environments and persistent dynamics is a possible direction for future work. 
Furthermore, while we focused on attractive memory footprints ($\omega > 0$), the navigation of cells and active particles often relies on self-avoidance~\cite{Troger2024} or chemorepellent trails~\cite{ Hokmabad2022}. These scenarios can be readily captured within our framework by allowing negative values of the memory response coefficient ($\omega < 0$).
\\ 
From an experimental standpoint, while existing studies have documented self-generated footprints (such as laminin or fibronectin deposition by MDCK cells ~\cite{d’Alessandro2021} or Psl exopolysaccharides by \textit{Pseudomonas aeruginosa}~\cite{Zhao2013}), direct quantification of memory-mediated pairwise interactions remains scarce. Our framework provides testable predictions: pair trajectories of interacting cells can be directly mapped using traction force microscopy or microfluidic track assays to quantify effective interactions. Emergent nonreciprocity can be experimentally engineered and studied by pairing wild-type cells with mutant or drug-treated populations exhibiting altered matrix deposition or integrin/Rac1 signaling pathways~\cite{Pankov2000}. Measuring how density fluctuations scale with single-cell parameters in multicellular assays would provide a direct test of our  theoretical framework.

\section{Acknowledgements}

J.P.B. acknowledges the  High Performance Computing Facility at NCBS for computational resources.
R.M. acknowledges support from the ANRF, India, through PMECRG (project ANRF/ECRG/2024/002036/PMS).


\setcounter{secnumdepth}{2}
\appendix


\setcounter{figure}{0}
\renewcommand{\thefigure}{A\arabic{figure}}
\setcounter{table}{0}
\renewcommand{\thetable}{A\arabic{table}}

\section{Dynamical equation of memory}
\label{app:modelRationale}

Keeping in mind that the substrate may have a finite memory capacity, \textit{i.e.}, that the local environment cannot be remodeled endlessly, we assume that the deposited footprints ($M_i$) on the $i^{\text{th}}$ site grow linearly with the number of visitations ($n_i$) at first and eventually saturate to a finite value $M_{\text{max}}$ called the memory capacity. We take $M_{\text{max}}$ to be site-independent. We model the memory deposition by a Hill function~\cite{Goutelle2008},
\begin{equation}\label{appEq:HillFunction}
    M_i (n_i) = M_{\text{max}} \frac{n_i}{\alpha^{-1} + n_i},
\end{equation}
which shows both the initial linearity and eventual saturation.
Here, the memory deposition coefficient $\alpha$ is defined such that, for a small number of visits (\textit{i.e.}, at early times), $M_i \approx \alpha M_{\text{max}} n_i$; that is, larger $\alpha$ leads to a proportionally larger footprint deposition. 
Rearranging Eq.~\eqref{appEq:HillFunction} leads to
\begin{equation}\label{appEq:nFromHill}
n_i = \frac{M_i \alpha^{-1}}{M_{\text{max}} - M_i},    
\end{equation}
which can be used in 
\begin{equation}\label{appEq:dM/dn}
    \frac{dM_i}{dn_i} = \frac{\alpha^{-1} M_{\text{max}}}{\left( \alpha^{-1} + n_i\right)^2}
\end{equation}
to eliminate the visit number ($n_i$) from the r.h.s. To convert the memory dynamics from being dependent on the visitation number to being time-dependent, we take $\frac{dM_i}{dt} = \frac{dM_i}{dn_i} \frac{dn_i}{dt}$, with $\frac{dn_i}{dt} = 1$ while a cell occupies site $i$. Allowing more than one cell to deposit memory on the same site at a time yields the first (deposition) term in the r.h.s. of Eq.~\eqref{eq:memory} of the main text.

\section{Velocity in external memory gradient}
\label{app:mnemotaxisDerivation}

Consider a cell without the ability to deposit memory (\emph{i.e.}, $\alpha = 0$) at the $i^{\text{th}}$ site on a lattice with a uniform gradient of pre-deposited footprint. The memory content at its current site is $M_i$ and $\Delta M = g \Delta x$, where $g$ is the uniform gradient and $\Delta x$ is the lattice spacing taken to be unity. Hence, the memory content at the $(i+1)^{\text{th}}$ and the $(i-1)^{\text{th}}$ sites are $M_{i \pm 1} = M_i \pm \Delta M$.

The velocity of the cell (toward the right) can be written as
\begin{align}
    v &= p_{i+1} - p_{i-1} \nonumber \\
    &= \frac{e^{\omega M_{i+1}} - e^{\omega M_{i-1}}}{e^{\omega M_{i+1}} + e^{\omega M_{i-1}}} \nonumber \\
    &= \frac{e^{\omega M_i} \left( e^{\omega \Delta M} - e^{-\omega \Delta M}\right)}{e^{\omega M_i} \left( e^{\omega \Delta M} + e^{-\omega \Delta M}\right)} \nonumber \\
    &= \tanh{(\omega \Delta M)}
\end{align}
 which appears in Eq.~\eqref{eq:velocity} of the main text. Note that both the lattice spacing and time-step are taken to be unity. 

\section{Effective interactions for $\epsilon > 0$}
\label{app:twoBodyFiniteDecay}

\begin{figure}[htbp]
    \centering
    \includegraphics[width=\columnwidth]{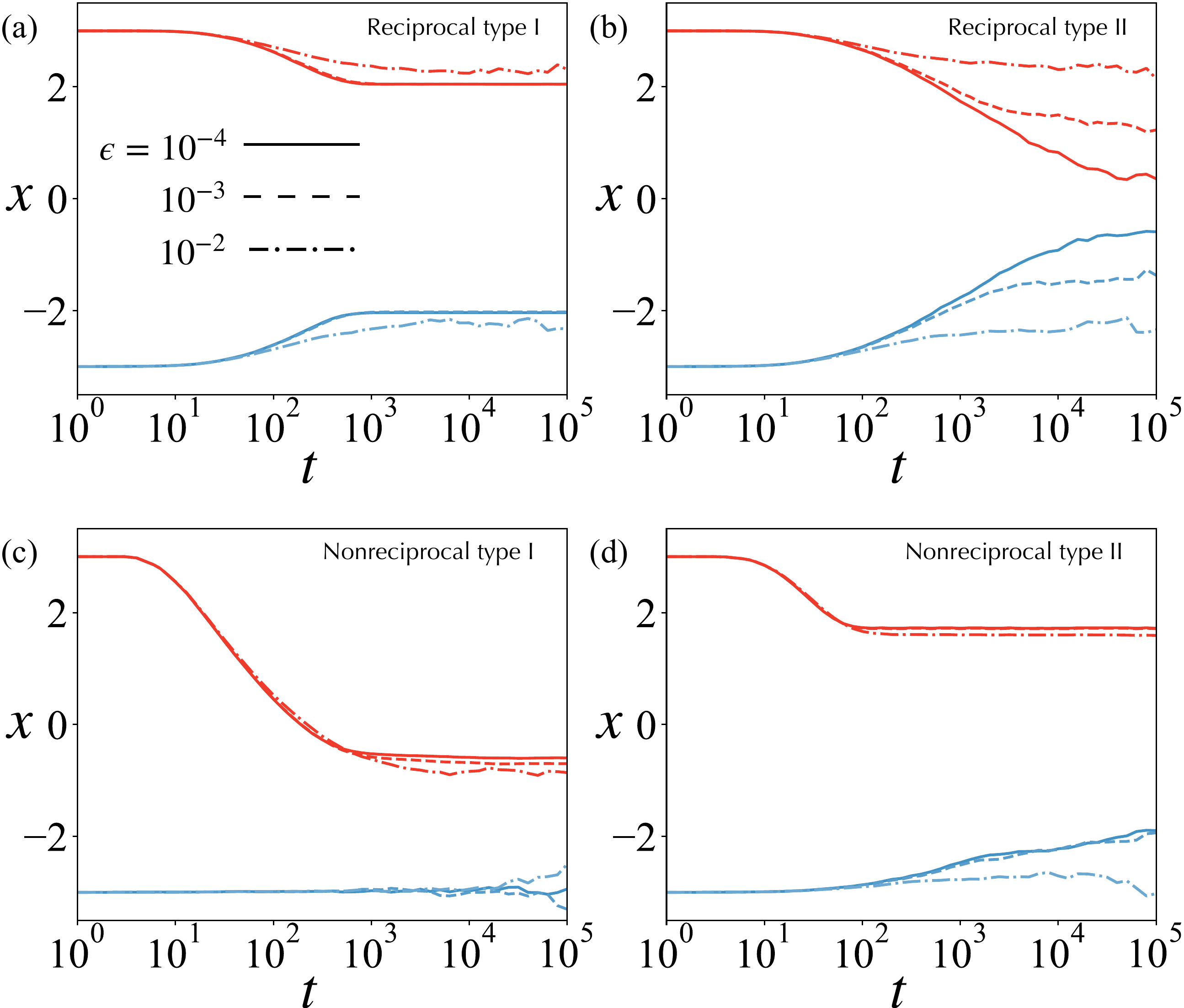}
    \caption{\textbf{Effect of finite memory decay rates ($\epsilon > 0$) on effective pairwise interactions.} Ensemble-averaged trajectories of two cells (red and blue curves) initiated at separation $x_0 = 6$ for three memory degradation rates: $\epsilon = 10^{-4}$ (solid lines), $\epsilon = 10^{-3}$ (dashed lines), and $\epsilon = 10^{-2}$ (dash-dotted lines). The four panels correspond to the interaction regimes shown in Fig.~5(a)--(d) of the main text: (a)~Reciprocal type~I, (b)~Reciprocal type~II, (c)~Nonreciprocal type~I, and (d)~Nonreciprocal type~II. All other model parameters ($\alpha_i, \omega_i$) in (a)--(d) are identical to those in Fig.~\ref{fig:5}(a)--(d), respectively.}
    \label{fig:AppC}
\end{figure}

In the main text (see Fig.~\ref{fig:5}(a)--(d)), we analyzed the emergent pairwise interactions between two cells in the absence of footprint degradation ($\epsilon = 0$). Here, we examine how a finite memory degradation rate ($\epsilon > 0$) modifies the ensemble-averaged two-cell trajectories across the four interaction regimes: Reciprocal type~I, Reciprocal type~II, Nonreciprocal type~I, and Nonreciprocal type~II (Fig.~\ref{fig:AppC}).

As shown in Fig.~\ref{fig:AppC}, at short to intermediate times ($t \ll \epsilon^{-1}$), memory deposition dominates over decay, and the trajectories for small degradation rates ($\epsilon = 10^{-4}$ and $10^{-3}$) closely follow the $\epsilon = 0$ behavior reported in Fig.~\ref{fig:5}(a)--(d). Once the individual footprint zones merge, the cells interact via the shared memory field and approach a quasi-stationary separation. However, at timescales comparable to or longer than the characteristic memory lifetime ($t \gtrsim \epsilon^{-1}$), footprint degradation erodes the shared memory field that sustains both self-caging and intercellular attraction. 

Specifically, for slow degradation ($\epsilon = 10^{-4}$, solid lines), the effective interactions remain nearly unaltered up to $t \sim 10^4$, showing only a mild relaxation at $t \sim 10^5$. For moderate degradation ($\epsilon = 10^{-3}$, dashed lines), the cells initially attract (or exhibit nonreciprocal pursuit in panels~(c) and~(d)), but as the deposited footprint degrades at $t \gtrsim 10^3$, the confining memory gradient weakens and the trajectories gradually drift apart. Finally, for rapid degradation ($\epsilon = 10^{-2}$, dash-dotted lines), footprints vanish before a persistent shared memory bridge can be established between the cells; consequently, the effective interaction is strongly suppressed and becomes transient across all four regimes.

\section{Effective interactions from trajectories}
\label{app:twoBodyEffectivePotentials}

To quantify the memory-mediated interactions between two cells, we extract effective pair potentials $V_{12}(r)$ and $V_{21}(r)$ from their ensemble-averaged trajectories. Note that, in general, these potentials also have a sense of direction. While nonreciprocal interactions cannot be derived from a single conservative Hamiltonian, one-dimensional directional potentials experienced individually by particles $1$ and $2$ characterize both the nature of the interaction and the degree of nonreciprocity~\cite{Shi2026}.

In the overdamped limit with unit friction ($\gamma = 1$), the effective forces equal the mean velocities and are computed via
\begin{equation}
    F_{12}(t) = \frac{dx_1(t)}{dt}, \,\, F_{21}(t) = \frac{dx_2(t)}{dt}
    \label{eq:force_velocity}
\end{equation}
where force profiles $F_{12}(r)$ and $F_{21}(r)$ are the forces on the first cell due to the second and on the second cell due to the first, respectively, obtained by expressing $F_{ij}(t)$ as a function of the separation $x_{\rm{rel}} = x_1(t) - x_2(t)$. We have $\partial/\partial x_1 = +d/dx_{\rm{rel}}$ and $\partial/\partial x_2 = -d/dx_{\rm{rel}}$, defining the directional potentials $V_{12}(x_{\rm{rel}})$ and $V_{21}(x_{\rm{rel}})$ via
\begin{equation}
    F_{12}(x_{\rm{rel}}) = -\frac{dV_{12}(x_{\rm{rel}})}{dx_{\rm{rel}}}, \qquad
    F_{21}(x_{\rm{rel}}) = +\frac{dV_{21}(x_{\rm{rel}})}{dx_{\rm{rel}}}.
    \label{eq:force_potential_def}
\end{equation}
Integrating Eq.~\eqref{eq:force_potential_def} with $V_{12}(l_{12}) = V_{21}(l_{21}) = 0$ gives
\begin{align}
    V_{12}(x_{\rm{rel}}) &= -\int_{x_0}^{x_{\rm{rel}}} F_{12}(x_{\rm{rel}}') dx_{\rm{rel}}' \nonumber \\
    V_{21}(x_{\rm{rel}}) &= \int_{x_0}^{x_{\rm{rel}}} F_{21}(x_{\rm{rel}}') dx_{\rm{rel}}'
\end{align}
For reciprocal interactions, we have $F_{21}(x_{\rm{rel}}) = -F_{12}(x_{\rm{rel}})$, $V_{12}(x_{\rm{rel}}) = V_{21}(x_{\rm{rel}})$ and $\dot{x}_{\mathrm{cm}} = 0$. However, in general, memory-induced asymmetry yields $F_{21}(x_{\rm{rel}}) \neq -F_{12}(x_{\rm{rel}})$, splitting the potentials $V_{12}(x_{\rm{rel}}) \neq V_{21}(x_{\rm{rel}})$ and driving net center-of-mass motion. The potentials obtained from Eq.~(D3) are then fitted with Eq.~\eqref{eq:potential} to extract $k_{ij}$, $l_{ij}$, and $\delta_{ij}$.

\section{Inferred interaction parameters}
\label{app:twoBodyInferredParam}

\definecolor{v12red}{HTML}{ef3b2c}
\definecolor{v21blue}{HTML}{4292c6}

\begin{table}[htbp]
\centering
\caption{Inferred fit parameters ($l_{ij}$, $k_{ij}$, and $\delta_{ij}$) of the effective directional potentials $V_{ij} = \frac{k_{ij}}{2}(x_{\mathrm{rel}} - l_{ij})^{\delta_{ij}}$ for reciprocal (R) and nonreciprocal (NR) interaction regimes.}
\label{tab:inferred_parameters}
\renewcommand{\arraystretch}{1.35} 
\setlength{\tabcolsep}{4.5pt}      
\begin{tabular}{|c|c|c|c|c|c|c|}
\hline
Regime & \textcolor{v12red}{$l_{12}$} & \textcolor{v21blue}{$l_{21}$} & \textcolor{v12red}{$k_{12}$} & \textcolor{v21blue}{$k_{21}$} & \textcolor{v12red}{$\delta_{12}$} & \textcolor{v21blue}{$\delta_{21}$} \\
\hline
R I   & \textcolor{v12red}{$4.09$} & \textcolor{v21blue}{$4.09$} & \textcolor{v12red}{$3.37 \times 10^{-3}$} & \textcolor{v21blue}{$3.32 \times 10^{-3}$} & \textcolor{v12red}{$1.96$} & \textcolor{v21blue}{$1.97$} \\
\hline
R II  & \textcolor{v12red}{$0.50$} & \textcolor{v21blue}{$0.50$} & \textcolor{v12red}{$1.69 \times 10^{-4}$} & \textcolor{v21blue}{$1.74 \times 10^{-4}$} & \textcolor{v12red}{$1.42$} & \textcolor{v21blue}{$1.42$} \\
\hline
NR I  & \textcolor{v12red}{$2.45$} & \textcolor{v21blue}{$2.45$} & \textcolor{v12red}{$6.65 \times 10^{-3}$} & \textcolor{v21blue}{$1.48 \times 10^{-4}$} & \textcolor{v12red}{$2.26$} & \textcolor{v21blue}{$0.92$} \\
\hline
NR II & \textcolor{v12red}{$4.58$} & \textcolor{v21blue}{$3.92$} & \textcolor{v12red}{$4.45 \times 10^{-2}$} & \textcolor{v21blue}{$1.41 \times 10^{-3}$} & \textcolor{v12red}{$2.16$} & \textcolor{v21blue}{$2.16$} \\
\hline
\end{tabular}
\end{table}


\bibliographystyle{apsrev4-2}
\bibliography{ref}

\clearpage

\end{document}